\documentclass[lnbip,sechang,a4paper]{svmultln}
\usepackage{amssymb}
\usepackage{graphicx}

\usepackage{subfigure}

\usepackage{makeidx}  
\begin{document}
\mainmatter              
\title{Towards Connected Enterprises: The Business Network System}
\titlerunning{The Business Network System}  
%
\author{Daniel Ritter}
%
%
\tocauthor{Daniel Ritter}
\institute{Technology Development -- Process and Network Integration, SAP AG, Dietmar-Hopp-Allee 16, 69190 Walldorf\\
\email{daniel.ritter@sap.com}}

\maketitle

\begin{abstract}
The discovery, representation and reconstruction of Business Networks (BN) from Network Mining (NM) raw data is a difficult problem for enterprises. This is due to huge amounts of complex business processes within and across enterprise boundaries, heterogeneous technology stacks, and fragmented data. To remain competitive, visibility into the enterprise and partner networks on different, interrelated abstraction levels is desirable.
We present a novel data discovery, mining and network inference system, called Business Network System (BNS), that reconstructs the BN - integration and business process networks - from raw data, hidden in the enterprises' landscapes. BNS provides a new, declarative foundation for gathering information, defining a network model, inferring the network and check its conformance to the real-world ``as-is'' network. The paper covers both the foundation and the key features of BNS, including its underlying technologies, its overall system architecture, and its most interesting capabilities.

\end{abstract}

\section{Introduction}
Enterprises are part of value chains consisting of business processes connecting intra- and inter-enterprise participants. The network that connects these participants with their technical, social and business relations is called a \emph{Business Network} (BN). Even though this network is very important for the enterprise, there are few - if any - people in the organization who understand this network as the relevant data is hidden in heterogeneous enterprise system landscapes. Yet simple questions about the network (e.g., which business processes require which interfaces, which integration artifacts are obsolete) remain difficult to answer, which makes the operation and lifecycle management like data migration, landscape optimization and evolution hard and more expensive increasing with the number of the systems. To change that, \emph{Network Mining} (NM) systems are used to discover and extract raw data \cite{lsna2012} - be it technical data (e.g., configurations of integration products like \emph{Enterprise Service Bus} (ESB) \cite{eip}) or business data (e.g., information about a supplier in a \emph{Supplier Relationship Management} (SRM) product). The task at hand is to provide a system, that automatically discovers and reconstructs the ``as-is'' BN from the incomplete, fragmented, cross-domain NM data and make it accessible for visualization and analysis.

Previous work on NM systems \cite{lsna2012} and their extension towards a holistic management of BN \cite{confenis2012} provide a comprehensive, theoretical foundation on how to build a system suited to this task. With the \emph{Business Network System} (BNS), we are exploiting this opportunity stated by these requirements. In particular, we are leveraging the work on the modeling and reconstruction of integration networks \cite{wlp2012}, conformance checking of implicit data models \cite{dqmst2012}, the client API \cite{icist2012} and the BN model \cite{bpmn2011} to deliver an emergent, holistic foundation for a declarative BN management system.

In this work we present a system for BN data management, that (semi-) automatically discovers and reconstructs the ``as-is'' BN and visualizes its interrelated perspectives on the network. The major contributions of this work are (1) a list of the most important requirements of a BNS, derived from previous work, and (2) an emergent enterprise-ready architecture for declarative BN data management from the existing NM components. The application of our inference approach to real-world enterprise landscapes and performance numbers of the query processing can be found in \cite{wlp2012} and \cite{icist2012}, hence are not part of this paper.

Section \ref{sec:bnm} guides from the theoretical work conducted in the area of NM \cite{lsna2012} and \emph{Business Network Management} (BNM) \cite{confenis2012} to the real-world requirements and capabilities of a BNS and sketches a high-level view on the system (refers to (1)). Section \ref{sec:nm} takes up the work on the inference \cite{wlp2012} and the BN model \cite{bpmn2011} to combine them towards a homogeneous approach with transformations from the various domain models in the customer landscapes up to the client API.  Based on that, a declarative programming approach is explained, which combines the model checking \cite{dqmst2012}, network reconstruction \cite{wlp2012} and data access approaches \cite{icist2012} towards the overall system architecture (refers to (2)). Section \ref{sec:relatedWork} reviews and discusses related work. Section \ref{sec:futureWork} concludes the paper and lists some of the future work.

\section{Building a System for Infering and Managing BNs} \label{sec:bnm}
The BN consists of a set of interrelated perspectives of domain networks (e.g., business process, integration, social), that provide a contextualized view on which business processes (i.e., business perspective) are currently running, implemented on which integration capabilities (i.e., integration perspective) and operated by whom (i.e., social perspective). To compute the BN \cite{wlp2012}, Network Mining (NM) systems automatically discover raw data from the enterprise landscapes \cite{lsna2012}. These conceptual foundations are extended to theoretically ground the new BN data management domain \cite{confenis2012}.

Based on this foundation and on related work, the fundamental requirements and capabilities of a BNS can be summarized to the following: \emph{REQ-1} The data for the reconstruction of the BN shall be (semi-)automatically discovered within the enterprise landscapes and cloud applications, only guided by domain-specific configurations from a domain expert (from \cite{lsna2012}), \emph{REQ-2} There shall be a common model, suitable for the network reconstruction task, to which the domain-specific data is transformed. The inference approach shall be independent from the domains (from \cite{wlp2012,processMining2011}), \emph{REQ-3} The transformed data shall be loaded to scalable and continuously running BN inference programs (from \cite{wlp2012,lsna2012}), \emph{REQ-4} The BN inference programs shall be described declaratively, that the source code must not be changed and non-technical personas (e.g., the domain expert) can flexibly specify own programs (from \cite{wlp2012}), \emph{REQ-5} and shall allow cross-domain and enterprise/ tenant reconstruction, which requires a system accessible by all enterprises in the network, e.g., a public or private cloud setup (from \cite{wlp2012}), \emph{REQ-6} It is assumed that the model represents the real world ``as-is" artifacts. However, the data comes possibly from different sources or user input and has to be checked for compliance before loading to the raw data store (from \cite{dqmst2012,processMining2011}), \emph{REQ-7} The system shall support different perspectives (i.e., views) on the BN (e.g., business process, integration) (from \cite{bpmn2011}), and \emph{REQ-8} The client API shall allow remote access as well as scalable query, traversal and full-text search across the interconnected BN perspectives (e.g., through index creation) (from \cite{icist2012}).

\begin{figure}
\centering
\label{fig:system_nm}
\includegraphics[width=\textwidth]{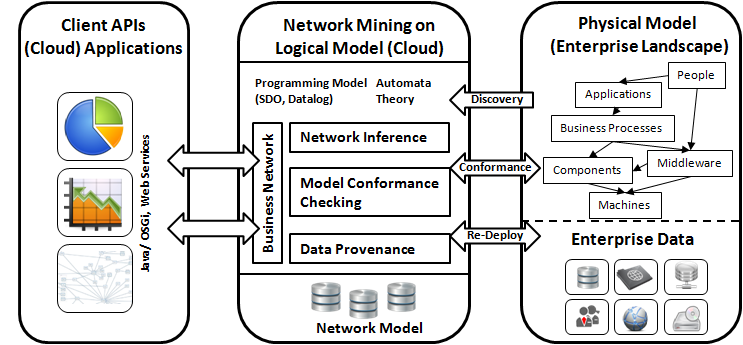}
\caption{Network Mining a la Business Network System}
\end{figure}

To sketch an idea on what these requirements mean for the construction of a BNS, Figure \ref{fig:system_nm} provides a high-level view on the core capabilities of our BNS. On the bottom right enterprise data is depicted, which is a mix of business process, social and integration artifacts stored in databases, packaged applications, system landscape directories (e.g., SAP SLD \cite{sapsld}), middleware systems (e.g., SAP PI \cite{sappi}), documents/ files, application back-ends, and so on. When pointed to an enterprise data source through configuration by a domain expert, the BNS introspects the source's metadata (e.g., WSDL file for Web service), discovers and transforms the domain data to a common representation (see \emph{REQ-1, REQ-2}). Other functional and queryable data sources are similarly processed. The common representation, referred to as the physical model in Figure \ref{fig:system_nm} (i.e., later called inference model), is a uniform, formalization of the enterprise's data sources (see \emph{REQ-2}).
 The center of Figure \ref{fig:system_nm} shows the core elements of an NM system, theoretically discussed in \cite{lsna2012,confenis2012}, which computes the perspectives of the BN for client access. Rather than interacting directly with the underlying data sources, or even with the more uniform physical model, a BNS client application works on the BN perspectives (i.e., the logical model, later called BN model). For that, the physical model is loaded and checked for conformance to the inference model by a declarative model checker based on automata theory \cite{dqmst2012} (see \emph{REQ-6}). After the loaded data has been checked it is stored as raw data for the continuously running network reconstruction programs (see \emph{REQ-3}) using logic programming (i.e., our approach uses Datalog due to rationale discussed in \cite{wlp2012}; see \emph{REQ-4}). Since the BN reconstruction works on cross-domain and enterprise data, and (cloud) applications want to access the BN data, the NM-part of the system is located in the public or private cloud (see \emph{REQ-5}), while the discovery-part is located in the enterprise system landscapes. This results in interrelated network perspectives, which are accessed by the clients for network visualization, simulation or analytics (see \emph{REQ-7}).

\section{Network Mining in the BNS} \label{sec:nm}
\subsection{The BNS Models}
\begin{figure}
\centering
\includegraphics[width=\textwidth]{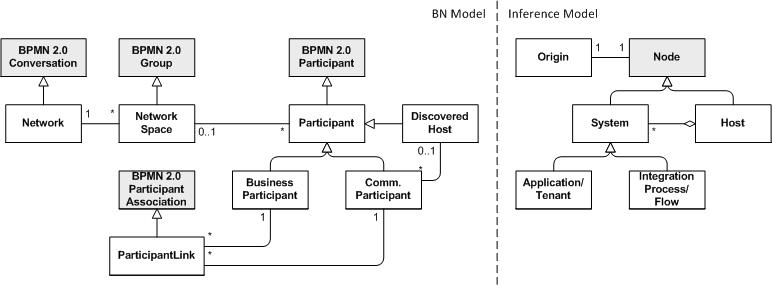}
\caption{Excerpts from the inference model (right) and BN model (left) showing only parts of the node definition.}
\label{fig:models}
\end{figure}

The premise of NM is that all information required to compute the BN is available in the enterprises' landscapes. In fact, the information can be found scattered across divers data sources, which come with different meta-data, formats, data semantics and quality (e.g., a SAP PI Business System \cite{sappi} represents a node in the BN, later called participant or system. The information about its interfaces can be found in the middleware system, e.g., SAP PI, however its description and physical host it runs on is usually maintained in a system landscape directory, e.g., SAP SLD \cite{sapsld}). The network inference approach must not know about the domain-specificities, but should be more generally able to identify an entity and assign all its data fragments (see \emph{REQ-2}). Hence, the BNS provides deployable components specific to the different domains, which can be configured to extract information from the data sources, pre-analyzes and transforms it into the inference model, defined in \cite{wlp2012} (see \emph{REQ-1}). Since in our approach the inference programs are written as Datalog rules \cite{datalog}, the inference model is represented as Datalog facts. Figure \ref{fig:models} (right) shows the node part of the inference model, whose basic entities are \emph{system} and \emph{host}. Each entity in the inference model (e.g., \emph{node}) holds a reference to its origin (i.e., meta-data about the data source and original object instance).

The distributed, domain-specific analysis decentralizes the information discovery process and guarantees that no domain logic has to be encoded in the inference programs (see \emph{REQ-2}). The decentralized discovery components collect the transformed information and push them to the inference system at a configurable schedule. With this operation, the current state of the system is moved to the inference system, which stores the information in the local knowledge base as raw data. Through that, the inference programs only rely on local raw data, which ensures more efficient processing and makes it less dependent on network communication. However, the attached \emph{origin} information keeps the link to the original records in the respective data sources (i.e., needed for provenance and continuous loading from the same source).

In Datalog, the raw data represents the extensible database (EDB) relations, which are evaluated by intensional database (IDB) rules (i.e., the inference programs). The result of the evaluation is the BN, which is already represented in a computer readable, visualizable standard as network-centric BPMN \cite{bpmn2011}. Figure \ref{fig:models} (left) shows the node part of the BN model, derived from the BPMN 2.0 collaboration-conversation diagram. The \emph{Participant} corresponds to the \emph{system} from the inference model. In addition, the inference programs compute \emph{ParticipantLink}, which represent semantic contextualization between participants from different network perspectives  (see \emph{REQ-6}). The business process and integration perspectives are built-in and defined as \emph{NetworkSpaces}, which are part of the \emph{Network} (i.e., BN; see \emph{REQ-7}).

\subsection{Declarative Network Mining and Extensibility}
The choice for two models throughout the whole architecture (i.e., inference model and BN model) helps to separate the distributed discovery and inference from the (remote) access for search, query and traversal. Furthermore this allows for a possibly different model evolution between the two models (e.g., new fields, entities added to the BN do not necessarily need to be propagated to the inference model). That means, only information from the automatic discovery has to be added in the inference model and programs. More concrete, an end-to-end model extension would require the (1) inference model and (2) its conformance checks, (3) the BN model, (4) the inference programs, and (5) the indices for query and traversal to change. Clearly, this is no task that should be done manually in the source code. Rather a configurable, declarative, model-centric approach is preferable (see \emph{REQ-4}).
\begin{figure}
\centering
\includegraphics[width=\textwidth]{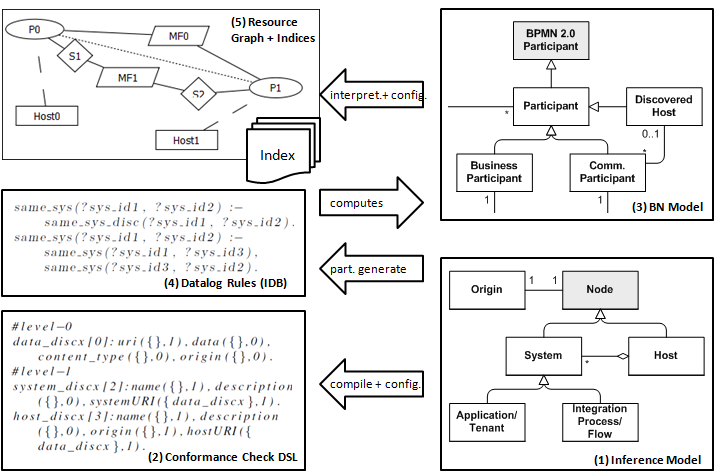}
\caption{Declaration and configuration capabilities of the BNS}
\label{fig:declarative}
\end{figure}

Figure \ref{fig:declarative} shows the end-to-end declaration and configuration approach chosen for the BNS. The inference model provides the specification for the compilation and configuration of the conformance checks (details on the DSL are in \cite{dqmst2012} and the Datalog programs for the inference task are in \cite{wlp2012}). While the conformance checks can be generated and configured without additional manual adaptation (see \emph{REQ-6}), the inference programs can only partially be generated, since they represent domain expert knowledge (e.g., business process, integration), which cannot be derived from the inference model yet. However, this is a natural extension point, which allows domain experts to write their own inference programs to tweak the BN (see \emph{REQ-4}). The BN model has to be adjusted before to allow the Datalog program to reference the modified entities. The resource graph representation and index structures for search, query and traversal of the BN adapt to changes in the BN model through a generic model interpretation (see \emph{REQ-8}).

The remaining major, manual effort are the mapping of the domain-specific artifacts to the modified inference model and the adaptation of the source code that discovers the information in the enterprise landscape. Since these tasks still require significant expert knowledge and vary between enterprise landscapes, the automation is left for further research.

\subsection{Continuous Business Network Reconstruction}
The raw data is pushed from the discovery components (i.e., from different enterprise landscapes) into the knowledge base in the BNS (i.e., cloud system). Since the records in the knowledge base are Datalog facts and the inference programs are represented as Datalog rules, a Datalog system runs the programs without additional transformations. For each of the defined network perspectives in the BN, a set of Datalog rules is defined and executed during the inference task.

For instance, the programs for the integration perspective have to cope with challenges like (a) cross-middleware inference, (b) combination of embedded and mediated communication, (c) message flow reconstruction (i.e., edges), and (d) fragmented information registered in different systems and with different semantics (e.g., runtime, system landscape, and configuration data). The basic entities relevant for the inference are nodes, i.e., logical entities like applications or tenants, called \emph{System} (see Figure \ref{fig:models} (right)), running on physical hosts, and edges representing the communication between systems via messages, the \emph{MessageFlow} (not shown). Technically, messages are exchanged over interfaces, \emph{Interface}, and channels, containing e.g. services, bindings and operations, which we represent as \emph{IncomingConfiguration} and \emph{OutgoingConfiguration} (not shown). The inbound and outbound configurations are considered separate entities, since they carry important information about the message flows, thus helping to reconstruct the edges. 
\begin{figure*}[!ht]
\begin{center}$
\begin{array}{cc}
\subfigure[System and Host facts]{\label{fig:sample_facts}\includegraphics[width=0.25\textwidth]{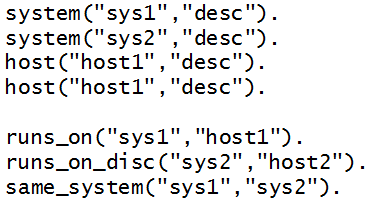}} &
\subfigure[Abstract sample rule on Host equivalences]{\label{fig:sample_host_equivalences}\includegraphics[width=0.35\textwidth]{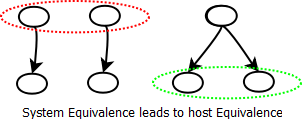}} 
\subfigure[Executable inference plan for Host equivalences]{\label{fig:host_exec_plan}\includegraphics[width=0.40\textwidth]{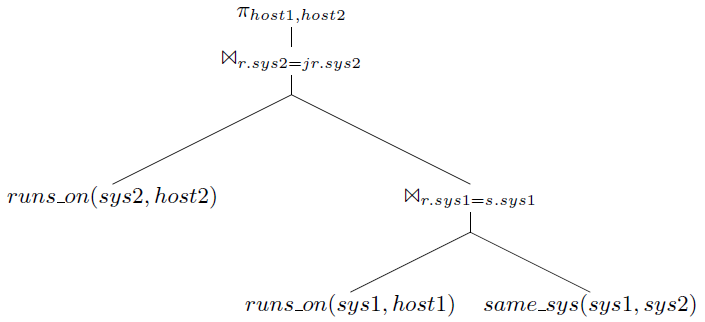}}
\end{array}$
\end{center}
\caption{Sample Datalog facts showing how a detected system equivalence leads to a host equivalence (with corresponding Datalog execution plan)}
\end{figure*}
Fig. \ref{fig:sample_facts} shows sample \emph{System} and \emph{Host} data as well as additional knowledge about their relationship. This data is represented as Datalog facts according to the inference model. Semantic relations between e.g. \emph{System} and \emph{Host} entities are \emph{runs\_on} as ``implemented-by" semantic (see Figure \ref{fig:sample_facts}), further specify the systems (refers to challenge (d)).
\begin{figure*}[!ht]
\begin{center}$
\begin{array}{cc}
\subfigure[Outbound/ Inbound call graph for deriving Message Flows]{\label{fig:cal_graph_flows}\includegraphics[width=0.5\textwidth]{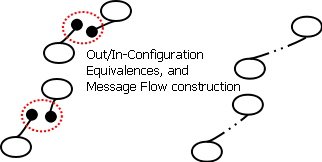}} &
\subfigure[Finding and merging equivalent systems]{\label{fig:call_graph_systems}\includegraphics[width=0.5\textwidth]{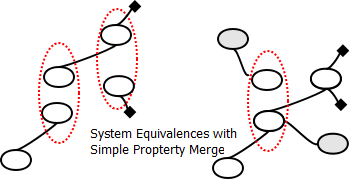}}
\end{array}$
\end{center}
\caption{Call graph showing message flow reconstruction and property merges}
\end{figure*}

\begin{figure*}
\centering
\includegraphics[width=\textwidth]{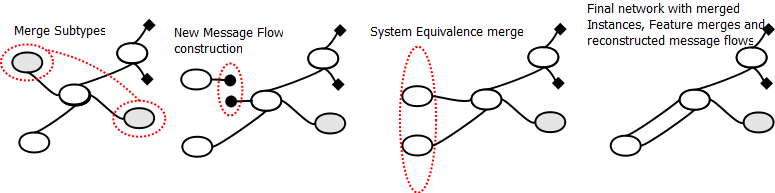}
\caption{Call graph showing complex entity property merge, message flow reconstruction and system equivalence merge}
\label{fig:stripe3}
\end{figure*}

For that, our inference approach exploits the knowledge about the basic entities to find instance equivalences, message flows and merge them to a network. the basic idea is to identify single, unique systems and compute call graphs on outgoing and incoming configurations from which message flows are derived (refers to challenge (c)). The single system call graphs are then merged to a multi-system graph, which denotes the integration network. Consider Fig. \ref{fig:cal_graph_flows}, which shows the call graph from configuration equivalences (dotted red circles around black configuration points) are computed leading to connected systems (circles). That leads to the construction of message flows and is usually done after \emph{System} equivalences (dotted red circle) are found as depicted in Fig. \ref{fig:call_graph_systems}. The simple (diamonds) and complex (grey circles) system properties are checked for merge (refers to challenges (a,b,d)). In this case the simple properties are added to the joint system instance and the complex properties are equivalent, thus merged to one instance (see Fig. \ref{fig:stripe3}, ``Merge Subtypes").

During the continuous process of unique instance detection (see \emph{REQ-3}), flow reconstruction and call graph merge, new instances from new data sources can be added (see Fig. \ref{fig:stripe3}, ``New Message Flow construction"). For that, flows are reconstructed based on the outgoing and incoming call graphs and system equivalence is checked. These operations can be applied in any order. Fig. \ref{fig:stripe3}, ``Final Network" shows the resulting network after call graph merges. Since the new system is equivalent to the existing one, but the message flows are different, the systems are merged and connected to the rest of the network through two distinct flows. In addition, programs for cross perspective correlation strive to find (semantic) connections between participants (i.e., \emph{ParticipantLink}) and between message flows (i.e., \emph{MessageFlowLinks} derived from the BPMN 2.0 \emph{MessageFlowAssociation}; see Figure \ref{fig:models} (left)). The whole network inference process is designed to run continuously. In that way, the source data can be brought to the knowledge base as raw data in any order and at any time. The inference programs would then pick-up the current snapshot of the raw data, and thus guarantee that the data is eventually up-to-date and independent of the load order (see \emph{REQ-3}).

\section{Related Work} \label{sec:relatedWork}
Our approach for integration network representation and inference is based on Datalog, which is a well-researched topic \cite{datalog1978,datalog} that had its revival recently due to good parallelization capabilities, latest through \cite{boom2010,declarativeImperative}. Even in the enterprise analytics domain, Datalog was recently applied, mainly through work of \cite{logicblox2010}. However, these approaches address non-network inference domains for which they define extensions.
For the overall system approach, related work is conducted in the area of Process Mining (PM) \cite{processMining2011}, which sits between computational intelligence and data mining. It has similar requirements for data discovery, conformance and enhancement with respect to NM \cite{lsna2012}, but does not work with network models and inference. PM exclusively strives to derive BPM models from process logs. Hence PM complements BNM in the area of business process discovery.
Gaining insight into the network of physical and virtual nodes within enterprises is only addressed by the \emph{Host} entity in NIM, since it is not primarily relevant for visualizing and operating integration networks. This domain is mainly addressed by the IT service management \cite{itsm2006} and virtualization community \cite{xen2009}, which could be considered when introducing physical entities to our meta-model.
The linked (web) data research, shares similar approaches and methodologies, which have so far neglected linked data within enterprises and mainly focused on RDF-based approaches \cite{linkedData2009}. Applications of Datalog in the area of linked data \cite{datalogWeb2010} and semantic web \cite{semanticweb2010} show that it is used in the inference domain, however not used for network inference.


\section{Discussion and Future Work} \label{sec:futureWork}
In this work, we present a reference implementation of a Business Network System based on the theory on Business Networks \cite{lsna2012} and Business Network Management \cite{confenis2012} by combining some of the work on conformance checking \cite{dqmst2012}, business network inference \cite{wlp2012} and a client API \cite{icist2012} into an emergent, enterprise-ready architecture. The architecture consitutes a holistic network data management platform, reaching from the information retrieval, network mining and inference, up to the BN.

The topics of declarative, automatic information retrieval and inference program generation require further investigation.


\begin{thebibliography}{33}

\bibitem{processMining2011} van der Aalst, W.: Process Mining: Discovery, Conformance and Enhancement of Business Processes, 2011.

\bibitem{boom2010} Alvaro, P., Condie, T., Conway, N., Elmeleegy, K., Hellerstein, J.M., Sears, R.C.: BOOMAnalytics: Exploring Data-centric, Declarative Programming for the Cloud. In: EuroSys, 2010.

\bibitem{logicblox2010} Aref, M.: Datalog for Enterprise Applications -- From Industrial Applications to Rese. Datalog 2.0 Workshop, Oxford, 2010.




\bibitem{linkedData2009} Bizer, C. Heath, T., Berners-Lee, T.: Linked Data -- The Story so Far. International Journal on Semantic Web and Information Systems, Volumn 5, Issue 3, p 1--22, Elsevier, 2009.


\bibitem{xen2009} Chowdhury, N.M.M.K., Boutaba, R.: Network virtualization: state of the art and research challenges. Communications Magazine, IEEE, 2009.


\bibitem{datalog1978} Gallaire, H., Minker, J. (Eds.): Logic and Data Bases. Symposium on Logic and Data Bases, Advances in Data Base Theory, Plenum Press, New York, 1978.

\bibitem{declarativeImperative} Hellerstein, J. M.: The Declarative Imperative -- Experiences and Conjectures in Distributed Logic. Technical Report, Berkeley, 2010.

\bibitem{eip} Hohpe, G., Woolf, B.: Enterprise Integration Patterns: Designing, Building, and Deploying Messaging Solutions. Addison-Wesley Longman, Amsterdam, 2003.




\bibitem{semanticweb2010} Motik, B.: Using Datalog on the Semantic Web. Datalog 2.0 Workshop, Oxford, 2010.

\bibitem{itsm2006} O'Neill, P., et al.: Topic Overview -- IT Service Management. Technical Report, Forrester Research, 2006.

\bibitem{datalogWeb2010} Polleres, A.: Using Datalog for Rule-Based Reasoning over Web Data: Challenges and Next Steps. Datalog 2.0 Workshop, Oxford, 2010.

\bibitem{bpmn2011} Ritter, D., Ackermann, J., Bhatt, A., Hoffmann, F. O.: Building a Business Graph System and Network integration Model based on BPMN. In: 3rd International Workshop on BPMN, Luzern, 2011.

\bibitem{lsna2012} Ritter, D.: From Network Mining to Large Scale Business Networks. International Workshop on Large Scale Network Analysis (WWW Companion), Lyon, 2012.


\bibitem{icist2012} Ritter, D.: The Business Graph Protocol. The 18th International Conference on Information and Software Technologies (ICIST), Kaunas, 2012.

\bibitem{confenis2012} Ritter, D.: Towards a Business Network Management. The 6th International Conference on Research and Practical Issues of Enterprise Information Systems (Confenis), Ghent, 2012.

\bibitem{dqmst2012} Ritter, D., Rupprich, S.: A Data Quality Approach to Conformance Checks for Business Network Models. The 6th IEEE International Conference on Semantic Computing (ICSC), Palermo, 2012.

\bibitem{wlp2012} Ritter, D.: A Logic Programming Approach for the Reconstruction of Integration Networks. The 26th Workshop on Logic Programming (WLP), Bonn, 2012.


\bibitem{sappi} SAP Process Integration (2012), \\
http://www.sap.com/germany/plattform/netweaver/components/pi/index.epx

\bibitem{sapsld} SAP System Landscape Directory (2012), \\
http://scn.sap.com/docs/DOC-8042

\bibitem{datalog} Ullman, J. D.: Principles of Database and Knowledge-Base Systems Volume I. Computer Science Press, 1988.



\end{thebibliography}
\end{document}